# Structural and Dynamical Mechanisms of a Naturally Occurring Variant of the Human Prion Protein in Preventing Prion Conversion[*]

Yiming Tang (唐一鸣), Yifei Yao (姚逸飞), and Guanghong Wei (韦广红)[†]

Department of Physics, State Key Laboratory of Surface Physics and Key Laboratory for Computational Physical Science (Ministry of Education), and Multiscale Research Institute of Complex Systems, Fudan University, Shanghai 200433, People's Republic of China.

**Keywords**: Prion protein, V127 variant, Molecular dynamics simulations, Dynamic network analysis.

**PACS**: 87.15.-v, 87.14.E-, 87.15.ap

**Abstract**
Prion diseases are associated with the misfolding of the normal helical cellular form of prion protein (PrP$^C$) into the beta-sheet-rich scrapie form (PrP$^{Sc}$) and the subsequent aggregation of PrP$^{Sc}$ into amyloid fibrils. Recent studies demonstrated that a naturally occurring variant V127 of human PrP$^C$ is intrinsically resistant to prion conversion and aggregation, and can completely prevent prion diseases. However, the underlying molecular mechanism remains elusive. Herein we perform multiple microsecond molecular dynamics simulations on both wildtype (WT) and V127 variant of human PrP$^C$ to understand at atomic level the protective effect of V127 variant. Our simulations show that G127V mutation not only increases the rigidity of the S2-H2 loop between strand-2 (S2) and helix-2 (H2), but also allosterically enhances the stability of the H2 C-terminal region. Interestingly, previous studies reported that animals with rigid S2-H2 loop usually do not develop prion diseases, and the increase in H2 C-terminal stability can prevent misfolding and oligomerization of prion protein. The allosteric paths from G/V127 to H2 C-terminal region are identified using dynamical network analyses. Moreover, community network analyses illustrate that G127V mutation enhances the global correlations and intra-molecular interactions of PrP, thus stabilizing the overall PrP$^C$ structure and inhibiting its conversion into PrP$^{Sc}$. This study provides mechanistic understanding of human V127 variant in preventing prion conversion which may be helpful for the rational design of potent anti-prion compounds.

---

[*] Project supported by the Key Program of the National Key Research and Development Program of China (Grant No. 2016YFA0501702) and the National Natural Science Foundation of China (Grant No. 11674065).
[†] Corresponding author. Email: ghwei@fudan.edu.cn

**Introduction**

The transmissible spongiform encephalopathies (TSEs), or prion diseases, are a group of fatal neurodegenerative disorders including scrapie in sheep,[1,2] chronic wasting disease in deer,[3] spongiform encephalopathy in bovine,[4] Creutzfeldt-Jakob disease (CJD),[5,6] fatal familial insomnia (FII),[7,8] and kuru[9,10] in human. The "prion hypothesis", in other words, the TSEs are caused by the misfolding and aggregation of prion proteins (PrP), has dominated the field since 1982.[11–16] This hypothesis has also been extended to cover other neurodegenerations such as Alzheimer's disease, Amyotrophic Lateral Sclerosis (ALS), and Parkinson's disease, which are respectively associated with the aggregation of beta amyloids, TDP-43, and α-synucleins.[17–23]

The key hallmark underlying the pathological misfolding and aggregation of PrPs is the conformational transition from a helix-rich conformer ($PrP^C$, in which C stands for cellular) to a sheet-rich pathogenic (or scrapie-like) conformation ($PrP^{Sc}$).[24–26] The $PrP^C$ exists as a glycoprotein composed of 209 residues including a highly disordered N-terminal segment and a globular folded region comprising residues 125~231. NMR[27–29] and X-ray[30] studies have shown that the folded domain of human $PrP^C$ contains three α-helices (residues 144-154, 173-194, and 200-228, denoted respectively H1, H2, and H3) and two short antiparallel β-strands (residues 128-131, and 161-164, denoted respectively S1 and S2) (Fig. 1a). Although numerous works have suggested that $PrP^C$ is related to a wide range of different cellular processes,[31–36] its biological function remains largely unknown. On the other hand, the $PrP^{Sc}$ isoform has characteristics of enriched β-sheet structures, proteases-resistance and a high propensity to aggregate into amyloid fibrils.[37–39] However, solving the atomic-level structure of $PrP^{Sc}$ is very challenging, possibly due to its insolubility, short survival time, and its fast aggregation into toxic fibrils.[40–42]

About 10% to 15% of human prion diseases are caused by mutations of the PrP.[43] These mutations lead to the spontaneous misfolding of the protein and the formation of neurotoxic protofibrils and fibrils. Examples of the more than twenty reported disease-related mutations include D178N,[44,45] Q217R,[46] and T188R,[47] related respectively to FFI, Gerstmann–Sträussler–Scheinker syndrome (GSS), and CJD. On the contrary, a few mutations (M129V polymorphism,[48] E219K polymorphism,[49,50] E200K,[51] and G127V) are reported to have protective effect on prion diseases. For example, Mead and coworkers reported a novel acquired prion disease resistance factor (G127V mutation) selected during the kuru epidemic in Papua New Guinea.[52] Asante et al. further performed transgenic mice experiments and showed that the V127 variant completely prevents prion diseases and that its mechanism is distinct from the M129V polymorphism.[53] In addition, Sabareesan and Udgaonkar showed that the G126V mutation in mouse PrP (equivalent to G127V in human PrP) can slow the initial fibril growth by extending the lag phase.[54] In spite of these experimental advances, the molecular mechanisms underlying the complete resistance of the V127 variant to prion disease remain mostly unknown.

Molecular dynamics (MD) simulations have been widely used to investigate the structural and dynamical properties of proteins.[55–57] In 2016, Yao et.al. reported the first computational study on PrP dimer and its V127 variant. Their 100-ns simulations showed that the constructed PrP dimer with G127V mutation is less stable than the WT dimer.[58] Recently, Lin et.al solved the NMR structures of monomeric WT PrP and the G127V mutant and found G127V to be less stable than WT. They also performed 100-ns MD simulations starting from their NMR structures and showed that S1 and S2 in G127V is less populated than those in WT.[59] However, due to the short time scales of their simulations, the atomic-level mechanism underlying the G127V-mutation-induced protective effect of PrP is not well understood. Herein, we have carried out six independent 2-μs all-atom MD simulations on WT PrP monomer and its V127 variant. Our simulations show that the G127V mutation stabilizes the PrP monomeric structure by greatly enhancing the structural rigidity of the S2-H2 loop and the stability of the C-terminal of H2. The increased rigidity of S2-H2 loop and the enhanced H2 stability in V127 variant may inhibit the structural conversion from human PrP$^C$ to PrP$^{Sc}$, thus prevent prion disease. Our hypothesis is supported by the fact that prions are poorly transmissible to animals with their PrP$^C$ carrying a rigid loop and a recent finding showing that stabilization of H2 of prion protein prevents its misfolding and oligomerization. Interaction analysis and optimal allosteric path analysis provide the atomic-level mechanism of G127V in the enhancement of S2-H2 loop rigidity and the H2 C-terminal stability. Community network analysis further show that G127V mutation also increase the global correlations of PrP.

**Results and Discussion**

**G127V mutation increases the S2-H2 loop rigidity and H2 C-terminal stability**

We have performed three individual 2-μs-long simulations for both the globular domain of WT PrP (comprising residues 125~228) and its G127V mutant. The initial structure of WT PrP is taken from an solution NMR structure,[27] while the G127V mutation is obtained by replacing the sidechain of G127 with that of Valine. For each system, we first calculate the time evolution of backbone root mean square deviations (RMSDs) with respect to the native WT NMR structure. The RMSD values of WT in the three simulations oscillate more largely than those of G127V. For example, RMSD values in WT-1 run experience a slow and continuous increase from 0.15 nm to 0.40 nm in the first 1.5 μs, followed by a sudden drop after which the RMSD values fluctuate around their equilibrium value of ~0.35 nm (Fig. S1a). In contrast, the RMSD values in G127V-1 run fluctuates around ~0.25 nm during the full period of the simulation (Fig. S1d). The probability density functions (PDF) of the WT and G127V RMSD values in four consecutive time windows are shown in Fig. 1b-c. A shift from small RMSD values to larger ones can be clearly seen in the PDF of WT RMSD (Fig. 1a), while a similar G127V RMSD distribution is observed for all four time windows (a main peak centering 0.24 nm and a satellite peak centering 0.33 nm). These results show that the V127 variant has a relatively higher structural stability than WT.

We further investigate the influence of G127V mutation on the structural rigidity of PrP by calculating the root mean square fluctuation (RMSF) of both WT and G127V systems. As shown in Fig. 1d, the RMSF values of almost all residues in G127V is lower than those of WT (except for a few residues on the S1-H1 loop). In WT system, the residues in the S2-H2 loop (region 1, residues 164~174) and the residues in the C-terminal of H2 and those in H2-H3 loop (region 2, residues 192~197) have much higher RMSF values than other non-terminal residues, indicating their higher flexibilities than other residues. The high flexibilities of these two regions are consistent with previous atomistic and coarse-grained simulation studies on WT PrP.[60–64] It is noted that the RMSF values in our work are much higher than those reported in the previous all-atom simulation studies, probably due to the much longer simulation time in our work than previous studies (2 ns[60], 10 ns[61], 1.8 ns[65], 80 ns[62], 50 ns[63], 100 ns[58,59]). To the best of our knowledge, our work is the first microsecond-scale all-atom MD simulation study on prion protein. However, the RMSF values of residues in these two regions of the G127V systems are remarkably lower than those of the WT systems. These findings demonstrate that the G127V mutation increases the rigidity of the S2-H2 loop, and enhances the stability of C-terminal of H2.

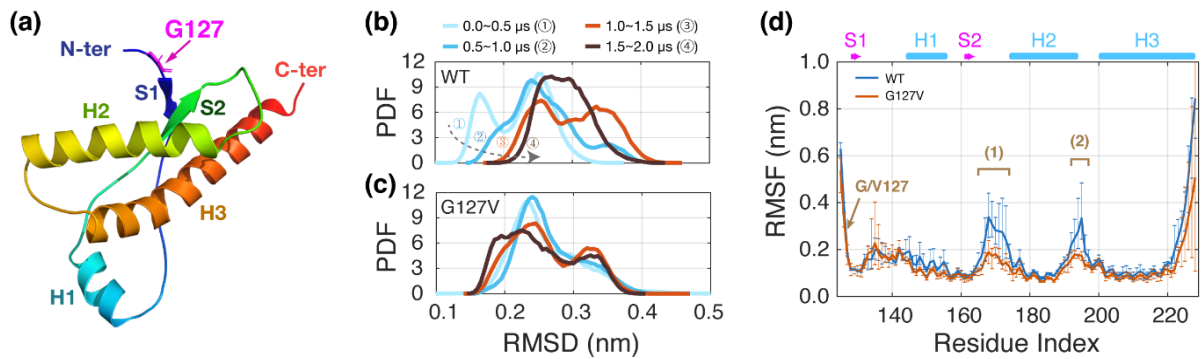

**Figure 1. Stability and rigidity of WT PrP and its V127 variant.** (a) A snapshot of the human PrP structure. (b-c) PDF of RMSD values in four consecutive time windows of (b) WT systems and (c) G127V systems. (d) RMSF of each residue on WT and G127V averaging over the three individual simulations for each system. The error bars are calculated by computing independent values from each individual simulation and taking the maximums and minimums of those values. The two regions where RMSF of G127V is remarkably lower than that of WT are labeled "(1)" and "(2)".

The misfolding of PrP monomer and the consequent oligomerization/fibrilization are closely related to the prion disease pathology.[26] The increased rigidity of S2-H2 loop as well as the enhanced stability of H2 C-terminal may stabilize the monomeric PrP$^C$ structure, thus inhibiting the prion conversion of human PrP$^C$ to PrP$^{Sc}$ and finally preventing prion disease. Our interpretation is well supported by the following experimental results. Firstly, the relationship between the rigidity of the S2-H2 loop and the PrP's resistance to prion diseases has been extensively discussed in previous studies.[66–69] Accumulating evidence shows that a transmission barrier exists between prion proteins carrying rigid S2-H2 loop and those carrying flexible one. And mice with a flexible S2-H2 loop is much more easily infected by prions than rigid S2-H2 loop mice.[70] Animals (such as bank vole, elk, horse and rabbit) with a rigid S2-H2 loop PrP usually do not develop prion diseases.[71–75] Secondly, the C-terminal region of H2 is identified as the first region to experience conformational conversion of PrP$^C$ to PrP$^{Sc}$ [76–78] and the stabilization of PrP H2 may prevent the misfolding and oligomerization of PrP.[79]

The solution-state NMR structures of bank vole, elk, and horse PrPs have been reported (PDBID: 2K56, 1XYW, 2KU4).[71,72,74] Fig. 2a shows the cartoon representations of their S2-H2 loops (residue 165~174). Surprisingly, they have very similar conformations, comprising a helical N-terminal (residue 165-168), and a turn-like region (residue 169-174), as highlighted respectively in magenta and blue boxes. The helical nature of residue 165-168 can be further characterized by calculating the (Phi, Psi) values of these residues. The (Phi, Psi) values of D167 are given in Fig. 2b, and they are distributed around (-80°, -10°), showing a propensity towards α-helical structure. Similar results are observed for residue M166 (Fig. S2).

To show whether the S2-H2 loop in V127 variant has a conformation similar to that in bank vole, elk, and horse PrPs, we perform cluster analysis on their conformations in the last 1 μs of the trajectories using a single-linkage algorithm[80] with an RMSD cutoff of 0.05 nm. As shown in Fig. 2c, the S2-H2 loop conformations in WT-1 run are clustered into 270 clusters, suggesting a wide variety of loop conformations. We show in Fig. 2f the representative configurations of the top four largest clusters. Interestingly, no helical N-terminal region is observed in the first and second cluster. Although the third and fourth clusters show a certain propensity towards forming helical-like structures, they only comprise respectively 7.6 % and 7.4 % of all conformations. Analyses of the other two WT systems give similar results (Fig. S3). In sharp contrast, for all of the three G127V MD runs, almost all S2-H2 loop conformations (containing respectively 97.5%, 99.7%, and 98.7% of conformations in G127V-1/2/3 system) are clustered into a single cluster, and all of them adopt a conformation similar to that of S2-H2 loop in bank vole, elk, and horse PrPs. We show in Fig. 2d-f the representative conformations of the S2-H2 loop in the largest cluster with the helical-like N-terminal region highlighted in magenta boxes. This propensity towards helical structure is further characterized by calculating the potential mean force (PMF) of residue D167 in WT and in G127V PrPs as a function of its (Phi, Psi) values. As shown in Fig. 2g, the PMF of WT D167 has three energy minimum basins. Two of them are located at (Phi, Psi) values of (-80°, -10°) and (-160°, -10°), while the third one is located at a (Phi, Psi) value of (-160°, ±180°). The deepest one at (-80°, -10°) corresponds to α-helical structures. In contrast, the D167 PMF of G127V system presents only two basins. The α-helical basin of G127V system is deeper and wider than that of WT system, suggesting that the S2-H2 loop of G127V system have a higher preference of forming helical N-terminal structures than that of WT. These results, together with the fact that prions are poorly transmissible to animals with their PrP$^C$ carrying a rigid loop, provide explanation to the strong resistance of V127 variant against prion diseases.

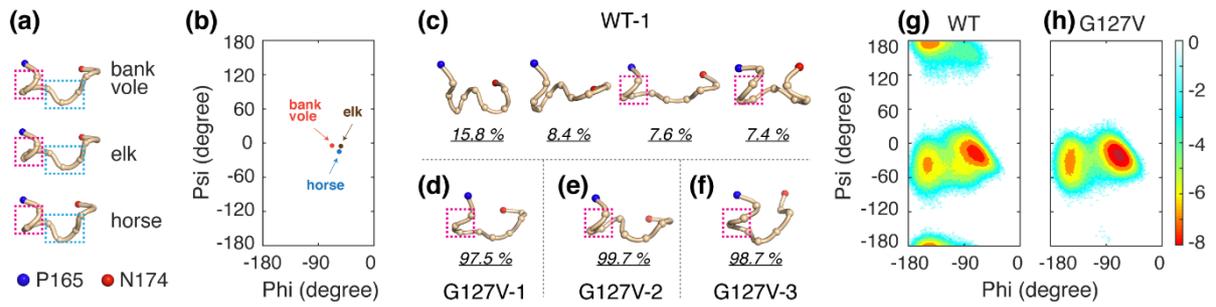

**Figure 2. Conformational characteristics of the S2-H2 loop.** (a) Snapshots of the S2-H2 loop (residue 165-174) in three prion-resistant PrPs. The red and blue dashed boxes correspond to the two structural characteristics: a helix-like structure and a turn-like loop. (b) The Ramachandran plot of residue D167 in bank vole, elk and horse PrPs. (c-f) Snapshots of the representative conformations of (c) the top four S2-H2 loop clusters in WT-1 MD run and of (c-e) the top one cluster in each G127V MD run. (g-h) PMF of D167 plotted as a function of the (Phi, Psi) values in (g) WT and (h) G127V PrPs.

**G127V mutation enhances local hydrophobic, H-bonding, and salt-bridge interactions in the vicinity of S2-H2 loop**

Although residues in the S2-H2 loop (residues 164~174) are distant from the mutation site (V127) in sequence, they are spatially close to each other (see Fig. 1a). We perform interaction analysis to understand how the G127V mutation improves the rigidity of S2-H2 loop. Valine has a bulky hydrophobic side chain while Glycine has no side chain, so we suspect that the hydrophobic interactions may play an important role. By calculating the contact probability between G/V127 and all residues in the S2-H2 loop, we find that the hydrophobic residue P165 has the highest contact probability with V127 (Fig. S5). The time evolution of P165-G/V127 contact number in WT-1 and G127V-1 runs show that the P165-G127 contact number fluctuates around 2 throughout the full period of simulation, while the P165-V127 contact number increases from ~2 to ~6 within the first 0.2 μs and fluctuates around 4 in the remaining simulation time (Fig. 3a). Similar result is seen from the other four simulations (WT-2/3, and G127V-2,3), and the contact number PDF calculated using a combined trajectory of the three individual simulations (Fig. S6b,c, Fig. 3b). These results suggest that the hydrophobic interactions between P165 and G/V127 is greatly enhanced as a result of the G127V mutation.

Interestingly, a recent work demonstrated that substituting G127 into I127 which possess higher hydrophobicity than both Glycine and Valine also significantly decreases the PrP cytotoxicity, indicating the important role of the V127-involved hydrophobic interaction in preventing prion diseases.[81]

The enhanced V127-P165 hydrophobic interaction may facilitate other residues in PrP fragment $_{125}$LGVYM$_{129}$ to interact with the S2-H2 loop ($_{162}$YYRPMDEY$_{169}$). We calculate the H-bond number between them. As shown in Fig. 3c, the H-bond number in G127V-1 run increases with simulation time and reaches ~5, while that in WT-1 run fluctuates around 3. Similar results are observed in WT-2/3 and G127V-2/3 systems (Fig. S6e-f), and a statistical analysis on all simulations for WT and G127V systems shows that G127V has a higher H-bond number than WT (Fig. 3d). A detailed analysis of the H-bonds show that residue E168 plays an important role on the H-bond formation: it forms stable H-bonds with three residues (L125, G126, and G/V127), and these H-bonds are much more stable in G127V system than in WT system (Fig. 3e). In addition, the M129-Y163 H-bond is stable in both WT and G127V system. The M129-Y163, and the E165-L125/G126/V127 H-bonds are shown in Fig. 3h-j. The enhancement of the E165-involved H-bonds may also contribute to the rigidity of the S2-H2 loop.

It has been reported that the R164-D178 salt bridge plays an important role on the structural stability of PrP.[82,83] To explore the effect of G127V mutation on the stability of this salt bridge, we calculate the distance between the charged groups of sidechains in R164 and D178 in WT-1 and G127V-1 run. Fig. 3f shows that the R164-D178 distance in both systems are close to ~0.4 nm in most of the simulation time, suggesting the existence of R164-D178 salt bridge in both systems. In addition, the distance in WT-1 run has a higher fluctuation than that in G127V-1. Analyses of the WT-2/3 and G127V-2/3 systems are given in Fig. S6h-I, and a statistical analysis using all three simulations is given in Fig. 3g. The PDF curve of R164-D178 distance in G127V systems has a single peak centering 0.38 nm, while that in WT systems has a dominant peak around 0.38 nm and a satellite peak centered at 0.45 nm. These data indicate that the R164-D178 salt bridge in G127V systems is slightly enhanced compared to that in WT.

Taken together, the G127V mutation enhances the V127-P165 hydrophobic interaction, the E165-L125/G126/V127 H-bonds, and the R164-D178 salt bridge. These interactions collectively rigidify the S2-H2 loop of PrP.

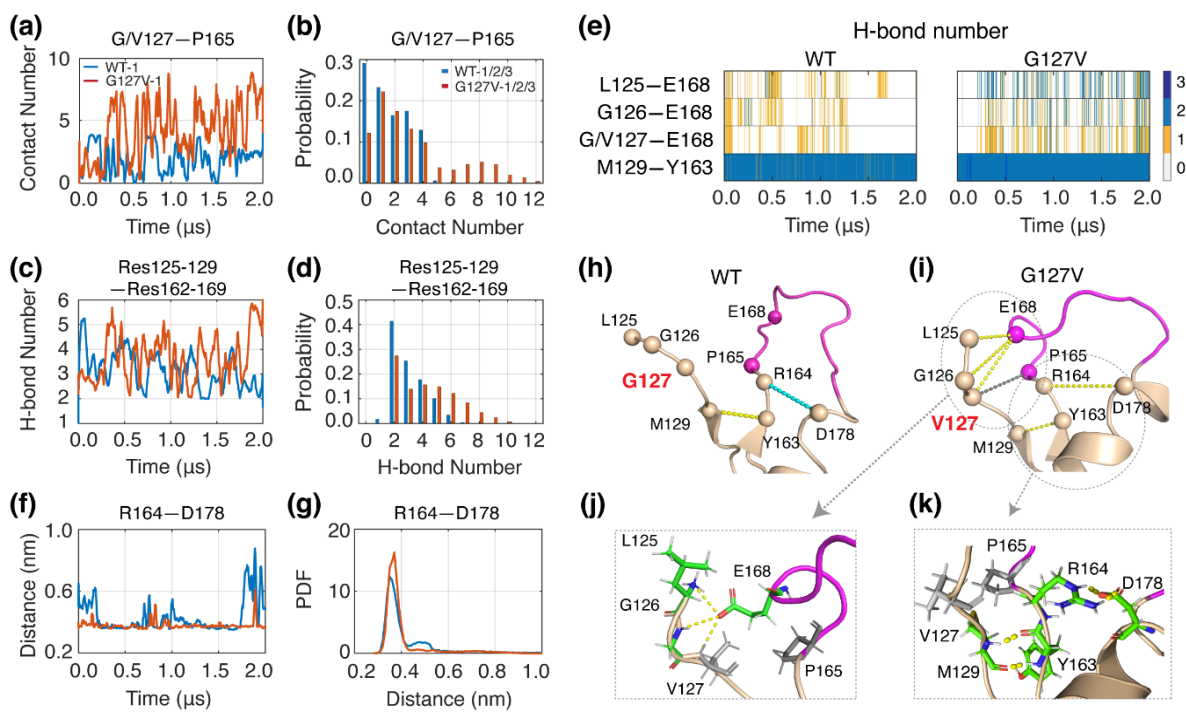

**Figure 3. Influence of G127V mutation on the interactions in the vicinity of residue 127 and S2-H2 loop.** (a,c,f) Time evolution of (a) contact number between G/V127 and P165, (c) number of H-bonds between residues 125~129 and residues 162~169, and (f) centroid distance between residue R164 and D178 charged sidechain groups, in WT-1 (blue line) and in G127V-1 (red line). (b,d,g) Statistical analysis using a combined trajectory of the last 1.0 μs in the three simulations of WT and G127V systems. (e) Time evolution of H-bond numbers between residues 125~129 and residues 162~169. Only residue pairs forming H-bonds in more than 1/4 of simulation time are shown. (h-i) Representative snapshots of the N-terminal region and the S2-H2 loop region of (h) WT system, and (i) G127V system. (j-k) Snapshots of the G127V showing (j) the E168-involved H-bonds, and (k) the R164-D178 salt bridge.

**Dynamical network analysis reveals the allosteric paths from the mutation site G/V127 to the C-terminal of H2**

As mentioned above, G127V mutation also enhances the stability of H2 C-terminal region (residues 192~197). This region is far from the mutation site (G/V127) both in sequence and in space, suggesting that G127V improves the H2 C-terminal stability through an allosteric effect. The optimal dynamical path analysis method has been demonstrated as a useful tool for identifying allosteric paths in various proteins.[84–88] Here we generate dynamic networks for WT and G127V systems by adding edges between residue pairs with a relatively high contact probability (>70%), and constructing the adjacency matrices using the absolute value of the inter-residue correlations (Fig. S4). We then calculate the optimal and suboptimal paths between residue G/V127 and G195 (the center residue in 192~197 region) to unravel how G127V mutation increases the stability of this distant region.

In WT PrP system, we identify two shortest network paths from G127 to G195 (path length = 317) as well as 15 paths with slightly longer path lengths (between 317 and 327). As shown in Table S1, all of these paths are similar: they start from G127, propagate to residue M129 on S1, and to V161 on S2, and subsequently to C terminus of H3, then pass to its N-terminal through nodes in H3 and finally reach residue G195. We present in Fig. 4a a snapshot showing residues located in the shortest path, and in Fig. 4b the correlation values of all edges alongside the path. As expected, all edges have high inter-residue correlation values (>0.6), and the edges within H3 have the highest correlation values, probably due to the existence of the stable H3 helix. Most of the edges (especially those located on the helices) connect two residues which are close in sequence, expect for the S1-S2 (M129-V161) and the S2-H3 (V161-V210) edges. These two edges are stabilized respectively by H-bonds between S1 and S2, and the V161-V210 hydrophobic interactions. To identify the important residues in this allosteric path, we remove the node corresponding to each residue and examine its effect on the optimal path length by calculating the path lengths after residue removal. As shown in Fig. 4c, removal of residue M129 results in the largest increase of optimal path (14%), revealing its highest importance in the G127-G195 allosteric path. In contrast, the optimal allosteric path from V127 to G195 in G127V system is much shorter than that in WT system (path length = 257). Similar to WT system, the V127-G195 allosteric paths start from V127, propagate to residue M129 on S1, and to Y162 on S2. However, after that, instead of propagating to C- terminal of H3, the allosteric paths propagate to the N-terminal of H2, and pass to its C-terminal through nodes in H2 and finally reach residue G195 (Table S2). We show in Fig. 4d all the residues on the shortest path, and in Fig. 4e the correlation values of all edges alongside the path in both WT and G127V systems. Except for the V127-M129 edge, most edges in G127V system have a higher correlation value than those in WT system, suggesting that the correlations in G127V system are much stronger than those in WT system. The main difference between the G127V allosteric path and the WT path is that S2 connects to H2 in G127V, while it connects to H3 in WT. As residues R164 and D178 are located respectively in S2 and H2, the S2-H2 correlation in G127V system may result from the enhanced R164-D178 salt bridge. The importance of S2-H2 correlation can be revealed by the alternation of optimal path length upon node removal. As shown in Fig. 4f, the removal of residue Y162 results in the largest increase in the optimal path length (4%). More interestingly, removing the S2-H2 interactions by cutting all edges connecting Y162 and H2 (Y162 - C192/I182/T183) leads to a suboptimal V127-G195 path (Fig. 4d and Table S3) that is similar to the G127-G195 optimal path in WT system, but with a shorter path length (282). These results provide insights into the allosteric effect by which G127V mutation enhances stability of the remote region (H2 C-terminal). The optimal allosteric path from G127 to H2 C-terminal in WT PrP passes edges connecting S2 and H3. After the G127V mutation, the interaction between S2 and H2 N-terminal is enhanced (through stabilized R164-D178 salt bridge), thus generates a new optimal allosteric path with a much shorter path length. This path passes through edges connecting S2 and H2. In addition, a path same to the WT optimal path also exists, but becomes the suboptimal path in G127V. The new-generated optimal path and the shortened suboptimal path from V127 to the G195 explains the increased long-range correlation between the mutation site and the C-terminal of H2, thus stabilize the structure of H2 C-terminal.

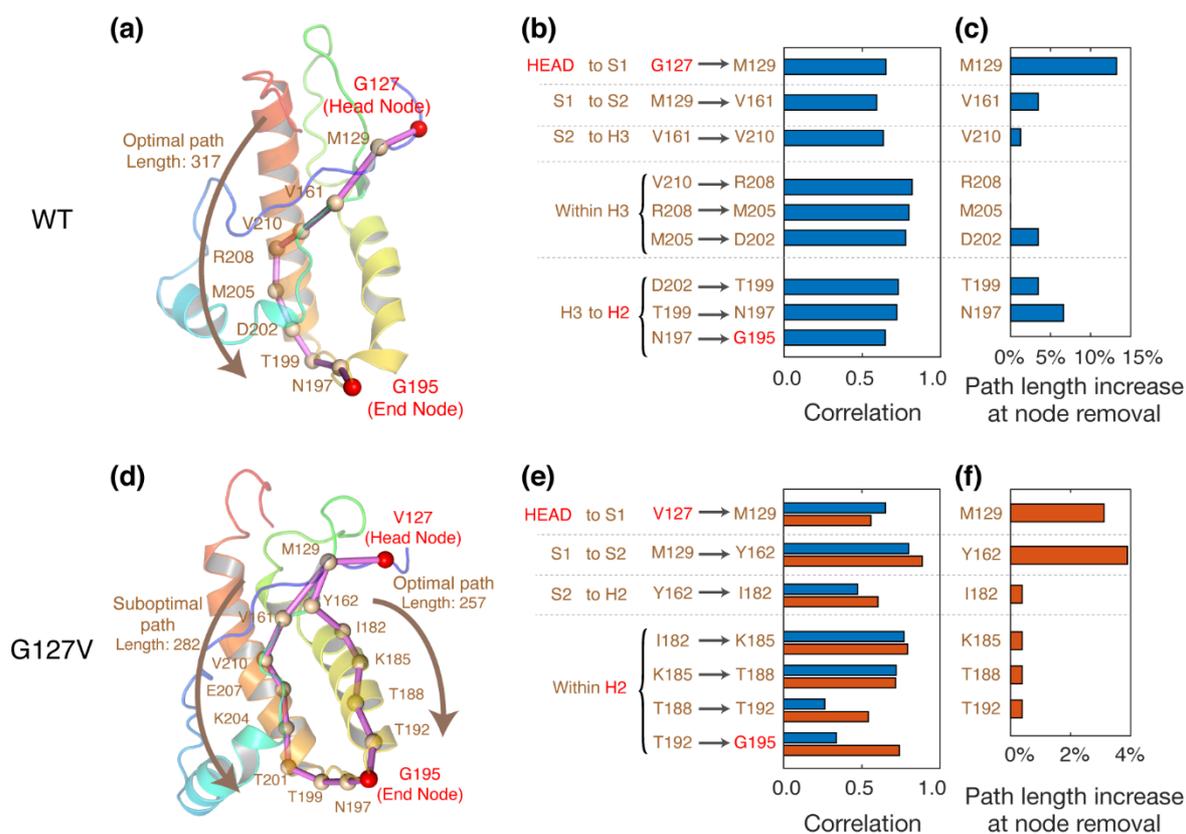

**Figure 4. Allosteric paths from the mutation site (G/V127) to the C-terminal of H2 in WT and G127V systems.** (a, d) Optimal path from residue G/V127 to G195 in (a) WT PrP and in

(d) G127V. (b, e) The correlation values of residue pairs forming the edges along the (b) G127-G195, and the (e) V127-G195 optimal path. (c, f) Percentage of optimal path length increase upon removal of each node of (c) WT and (f) G127V optimal paths.

**G127V mutation strengthens the global correlation of the PrP**

We further investigate the effect of G127V mutation on the global correlation of PrP by calculating the correlation values between each two residues. As shown in Fig. 5a, the WT PrP correlation map contains three diagonal positively-correlated regions corresponding to the inter-residue interactions in the three helices (red circles), an off-diagonal region corresponding to the anti-parallel H2-H3 interaction (blue circle), and two regions corresponding to H2/H3-S2 interactions (green circles). The G127V correlation map has a similar pattern to WT, but the correlation values of the six positively-correlated regions are mostly higher than those of WT (Fig. 5b). In addition, more positively-correlated regions are observed in the correlation map of G127V system. These results demonstrate that the inter-residue correlations in PrP significantly increase after G127V mutation.

Community network method has become a useful strategy for analyzing global correlations for biomolecules.[89–93] Based on the dynamical networks of WT and G127V, we calculated the optimal community distributions using the Girvan-Newman algorithm.[94] The residues belonging to the same community are more strongly and densely interconnected to each other, and have weaker connections to other residues in the protein. As shown in Fig. 5c, the WT community network comprises ten communities, including a single community for H1, two separate communities for N and C terminal of H2 (H2N and H2C), three separate communities for N-terminal, middle part, and C-terminal of H3 (H3N, H3M, and H3C), and one single community for S2-H2 loop. In sharp contrast, the G127V community network comprises only six communities. The three WT communities corresponding to S1/S2, H2N, and H3M have combined into one large community, and the two WT communities corresponding to H3N and H2C have also combined into one large community. In addition, the inter-community connectivity in G127V is stronger than that in WT (the width of edges connecting communities corresponds to the inter-community connectivity strength). The decrease in cluster number, the combination of several clusters into single large cluster, as well as the increase of inter-community connectivity strength indicate that the G127V mutation enhances the global correlations and interactions in PrP, thus stabilizes the PrP$^C$ structure and inhibits the prion conversion to PrP$^{Sc}$.

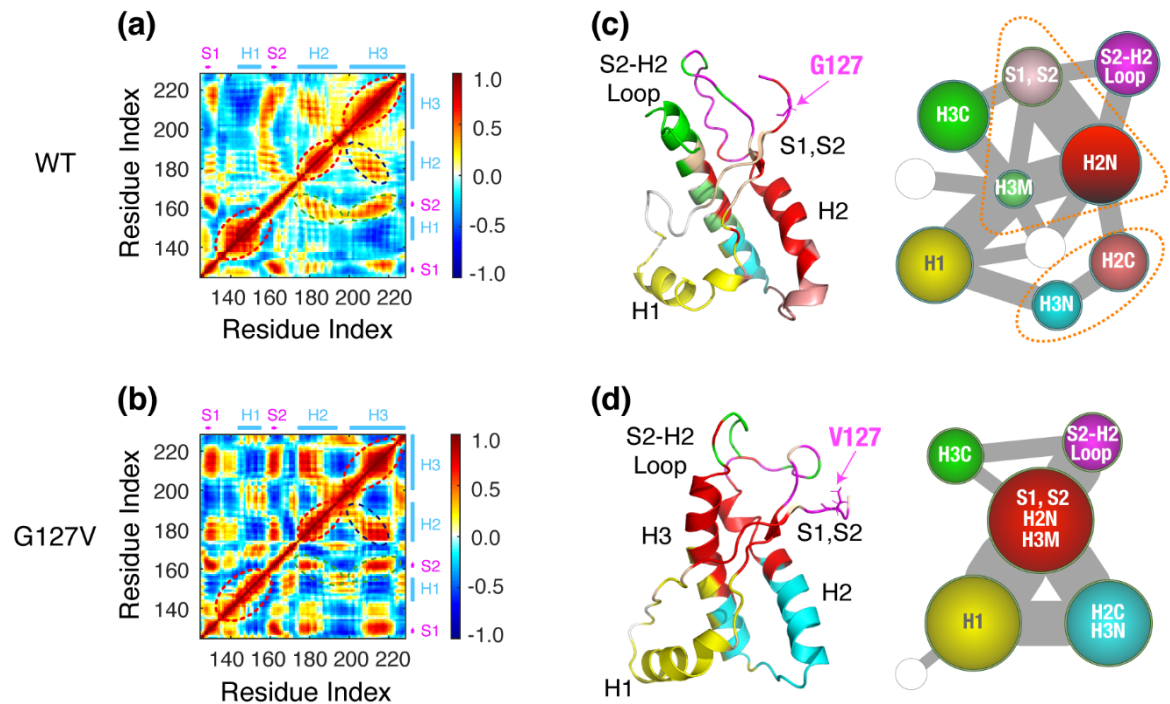

**Figure 5. Correlation and community network analysis of WT and V127 variant.** (a-b) Inter-residue correlation matrices of (a) WT and (b) G127V systems. (c-d) Community networks of (c) WT and (d) G127V systems. Left panels: snapshots of the proteins colored by communities. Right panels: schematic diagrams of the community networks. Each circle represents a single community. The size of the circle and the width of the edges correspond respectively to the size of the community and the connectivity strength between two communities.

**Conclusions**

We have performed six 2-μs-long MD simulations on the globular domain of WT human PrP and its V127 variant to investigate the effect of G127V mutation on the structural and dynamical properties of monomeric human PrP, with the aim of understanding the mechanism of G127V's disease-prevention. RMSD and RMSF analyses show that G127V mutation increases the structural stability and rigidity of PrP. Especially, (1) the rigidity of the S2-H2 loop is greatly increased, and (2) the stability of the C-terminal of H2 is enhanced. The increase rigidity and stability of these two regions may inhibit the prion conversion of PrP$^C$ into toxic PrP$^{Sc}$, thus prevents prion diseases. Our speculation is supported by the fact that prions are poorly transmissible to animals with rigid S2-H2 loop and recent findings showing that stabilization of H2 of prion protein prevents its misfolding and oligomerization. Interaction analysis show that the increased rigidity of the S2-H2 loop results from the enhanced V127-P165 hydrophobic interaction, E165-L125/G126/V127 H-bonds, and R164-D178 salt bridge. The mutation-induced stabilization of the R164-D178 salt-bridge enables the formation of a new allosteric path from mutation site to the H2 C-terminal with a much shorter path length than the optimal path in WT. The WT optimal path still exists in G127V and becomes the suboptimal path. The new optimal path and the shortened suboptimal path lead to an enhanced stability of the H2 C-terminal. At last, community network analysis shows that the global correlations and interactions in PrP is strengthened by the G127V mutation. Our findings provide structural and dynamical basis for understanding the role of human V127 variant in preventing prion conversion and propagation, which may be helpful for the rational design of potent anti-prion therapies.

## Material and Methods

### MD simulations

The initial structure of WT PrP globular domain was taken from a solution NMR structure comprising residues 125~228 (PDBID: 1HJM)[27]. We built the G127V structure by replacing the sidechain of Glycine with that of Valine using tools implemented in the Pymol package[95]. Three individual 2-μs-long MD simulations were performed for each system using the AMBER99SB-ILDN force field[96] in combination with the TIP-3P water model. All simulations were preformed using the GROMACS-5.1.2 package[97] in the NPT ensemble. Electrostatic interactions were treated using the Particle Mesh Ewald (PME) method[98] with a real space cut-off of 1.2 nm. The vdW interactions were calculated using a cutoff of 1.2 nm. The solute and solvent were separately coupled to an external temperature bath using a velocity rescaling method[99] and a pressure bath using the Parrinello-Rahman method[100]. The temperature and pressure were maintained at 310 K and 1 bar with a coupling constant of 0.1 ps, respectively. The neighbor-list was updated every 10 steps with a cut-off distance of 1.2 nm using a Verlet buffer[101].

### Trajectory analysis

Trajectory analysis was performed using our in-house-developed codes and the facilities implemented in the GROMACS-5.1.2 software package[97]. We discarded the first 1 μs data for each MD trajectory to remove the bias of the initial states. Therefore, the structural and dynamical properties of each system were based on the simulation data generated in the last 1 μs. The cluster analysis for the S2-H2 loop was performed using a single linkage method[80] with an RMSD cutoff of 0.05 nm. An inter-residue contact was defined when the aliphatic carbon atoms of two nonsequential side chains (or main chains) come within 0.54 nm or any other atoms of two nonsequential side chains (or main chains) lie within 0.46 nm. One H-bond is taken as formed if the N···O distance is less than 0.35 nm and the N–H···O angle is greater than 150°. Graphical analysis and visualizations were performed using the Pymol package[95].

**Generation of weighted dynamical networks**

Dynamical networks of WT and G127V were generated using our in-house-developed codes. Each $C_\alpha$ atom was selected as a representative node for the corresponding residue. Edges were added between two nodes if the corresponding residues are in contact during a majority of the simulation time (>70%). Nearest neighbors in sequence are not considered to be in contact as they lead to a number of trivial allosteric paths in the weighted network. The weight of each edge was defined as $-\log|C_{ij}|$, where $C_{ij}$ stands for the dynamic cross correlation of two nodes (i and j), so that the network distance between two nodes connected by an edge decreases as the correlation of two corresponding residues increases.

**Optimal and suboptimal allosteric path analysis**

On the basis of the dynamical networks, the allosteric signal transmission from mutation site (G/V127) to the C-terminal of H2 (residue 192~197) was analyzed by calculating the allosteric paths between residue G/V127 and residue G195. The length of a path $D_{ij}$ between distant residues i and j was defined as the sum of the edge weights between the consecutive nodes (k, l) along the path: $D_{ij} = \Sigma_{k,l} w_{kl}$. The path between residue i and j with the shortest distance $D_{ij}^{(0)}$ was found by using the Floyd-Warshall algorithm.[102] Because the high correlation values between two nodes may not necessarily guarantee strong physical interactions between the corresponding residues (it could also result from strong interaction between nearby residue pairs), the shortest path as well as the paths whose lengths lie within a certain limit of the shortest distance were equally important.[88] We defined the "optimal allosteric path" between residue i and j as the shortest paths between residue i and j (path length = $D_{ij}^{(0)}$) as well as the slightly longer paths (path length < $D_{ij}^{(0)} + 10$). The importance of each residue in the allosteric path was evaluated by calculating the change in optimal path length upon removing the node corresponding to this residue while keeping all other contacts in the network intact. Although the optimal allosteric path is the most dominant mode of communication between the two strongly-correlated remote residues, other paths with longer path length (corresponding to weak correlations/interactions) may also play a role. However, the optimal path analysis method [88] cannot identify these longer paths. Thus, we identified the most important residue in the allosteric effect by the aforementioned residue-removal technique, removed all edges connecting to this residue (these edges correspond to the strongest interaction along the optimal path), and calculated the path between residue i and j with the shortest distance $D_{ij}^{(1)}$ in this modified network. By removing these edges, the network path from the mutation site to its strongly-correlated remote residue has to pass through edges corresponding to weak interactions. The "suboptimal allosteric path" between residue i and j was thus defined as the path with path length $D_{ij}^{(1)}$ as well as the slightly longer paths (path length < $D_{ij}^{(1)}$ + 10). Identification of both optimal and suboptimal allosteric paths enables us to reveal the mechanism of the mutation-induced enhancement of global structural stability.

**Community network analysis**

We calculated the shortest network path between each residue pair (i, j, for i≠j). The betweenness of each edge was then defined as the number of shortest paths that cross that edge. The optimal community distribution was calculated using the Girvan-Newman algorithm[94], which iteratively removed the edge with the highest betweenness and recalculated the betweenness of all remaining edges until the modularity of the community network was maximized. The modularity is a measure of the quality of a particular division of a network, and the bigger the modularity, the better the division quality.[103] The optimal paths and dynamical networks were obtained using codes developed by Eargle and Sethi [88] with our modifications. Visualization of the community networks was performed using the VMD package[104] (snapshots) and the Omni Graffle package (schematic diagrams).

**Acknowledgements**

This work was supported by National Key Research and Development Program of China (Grant No. 2016YFA0501702) and the NSF of China (Grant No. 11674065). All simulations were performed using the high-performance computational facilities at the National High Performance Computing Center of Fudan University and the GPU Cluster in our group.

# References

[1]     GREIG J R 1950 Scrapie in sheep. *J. Comp. Pathol.* **60** 263–6

[2]     Hunter N 2003 Scrapie and experimental BSE in sheep *Br. Med. Bull.* **66** 171–83

[3]     Williams E S and Young S 1980 Chronic wasting disease of captive mule deer: a spongiform encephalopathy. *J. Wildl. Dis.* **16** 89–98

[4]     Chatigny M A and Prusiner S B 1980 Biohazards of Investigations on the Transmissible Spongiform Encephalopathies *Rev. Infect. Dis.* **2** 713–24

[5]     Will R G, Ironside J W, Zeidler M, Cousens S N, Estibeiro K, Alperovitch A, Poser S, Pocchiari M, Hofmar A and Smith P G 1996 A new variant of Creutzfeldt-Jakob disease in the UK *Lancet* **347** 921–5

[6]     Collinge J 1999 Variant Creutzfeldt-Jakob disease *Lancet* **354** 317–23

[7]     Lugaresi E, Medori R, Montagna P, Baruzzi A, Cortelli P, Lugaresi A, Tinuper P, Zucconi M and Gambetti P 1986 Fatal Familial Insomnia and Dysautonomia with Selective Degeneration of Thalamic Nuclei *N. Engl. J. Med.* **315** 997–1003

[8]     Medori R, Tritschler H-J, LeBlanc A, Villare F, Manetto V, Chen H Y, Xue R, Leal S, Montagna P, Cortelli P, Tinuper P, Avoni P, Mochi M, Baruzzi A, Hauw J J, Ott J, Lugaresi E, Autilio-Gambetti L and Gambetti P 1992 Fatal Familial Insomnia, a Prion Disease with a Mutation at Codon 178 of the Prion Protein Gene *N. Engl. J. Med.* **326** 444–9

[9]     Gajdusek D C 1963 Kuru *Trans. R. Soc. Trop. Med. Hyg.* **57** 151–69

[10]    Matthews J D, Glasse R and Lindenbaum S 1968 Kuru and cannibalism. *Lancet* **2** 449–52

[11]    Griffith J S 1967 Nature of the scrapie agent: Self-replication and scrapie *Nature* **215** 1043–4

[12]    Farquhar C F, Somerville R A and Bruce M E 1998 Straining the prion hypothesis *Nature* **391** 345–6

[13]    Zou W Q and Gambetti P 2005 From microbes to prions: The final proof of the prion hypothesis *Cell* **121** 155–7

[14]    Soto C 2011 Prion hypothesis: The end of the controversy? *Trends Biochem. Sci.* **36** 151–8

[15]    Prusiner S B 1982 Novel proteinaceous infectious particles cause scrapie *Science (80-. ).* **216** 136–44

[16]    Tuite M F and Serio T R 2010 The prion hypothesis: From biological anomaly to basic regulatory mechanism *Nat. Rev. Mol. Cell Biol.* **11** 823–33

[17]    Frost B and Diamond M I 2010 Prion-like mechanisms in neurodegenerative diseases *Nat. Rev. Neurosci.* **11** 155–9

[18]    Nussbaum J M, Schilling S, Cynis H, Silva A, Swanson E, Wangsanut T, Tayler K, Wiltgen B, Hatami A, Rönicke R, Reymann K, Hutter-Paier B, Alexandru A, Jagla W, Graubner S, Glabe C G, Demuth H U and Bloom G S 2012 Prion-like behaviour and tau-dependent cytotoxicity of pyroglutamylated amyloid-β *Nature* **485** 651–5

[19]    Chaari A and Ladjimi M 2019 Human islet amyloid polypeptide (hIAPP) aggregation in type 2 diabetes: Correlation between intrinsic physicochemical properties of hIAPP aggregates and their cytotoxicity *Int. J. Biol. Macromol.* **136** 57–65

[20]    Masuda-Suzukake M, Nonaka T, Hosokawa M, Oikawa T, Arai T, Akiyama H, Mann D M A and Hasegawa M 2013 Prion-like spreading of pathological α-synuclein in brain *Brain* **136** 1128–38

[21]    Chu Y and Kordower J H 2015 The Prion Hypothesis of Parkinson's Disease *Curr. Neurol. Neurosci. Rep.* **15** 1–10

[22]     Polymenidou M and Cleveland D W 2011 The seeds of neurodegeneration: Prion-like spreading in ALS *Cell* **147** 498–508

[23]     Prasad A, Bharathi V, Sivalingam V, Girdhar A and Patel B K 2019 Molecular mechanisms of TDP-43 misfolding and pathology in amyotrophic lateral sclerosis *Front. Mol. Neurosci.* **12** 25

[24]     Abid K and Soto C 2006 The intriguing prion disorders *Cell. Mol. Life Sci.* **63** 2342–51

[25]     Riesner D 2003 Biochemistry and structure of PrPC and PrPSc *Br. Med. Bull.* **66** 21–33

[26]     Kupfer L, Hinrichs W and Groschup M 2009 Prion Protein Misfolding *Curr. Mol. Med.* **9** 826–35

[27]     Calzolai L and Zahn R 2003 Influence of pH on NMR structure and stability of the human prion protein globular domain *J. Biol. Chem.* **278** 35592–6

[28]     Zahn R, Liu A, Lührs T, Riek R, Von Schroetter C, Garcia F L, Billeter M, Calzolai L, Wider G and Wüthrich K 2000 NMR solution structure of the human prion protein *Proc. Natl. Acad. Sci. U. S. A.* **97** 145–50

[29]     Calzolai L, Lysek D A, Güntert P, Von Schroetter C, Riek R, Zahn R and Wüthrich K 2000 NMR structures of three single-residue variants of the human prion protein *Proc. Natl. Acad. Sci. U. S. A.* **97** 8340–5

[30]     Knaus K J, Morillas M, Swietnicki W, Malone M, Surewicz W K and Yee V C 2001 Crystal structure of the human prion protein reveals a mechanism for oligomerization *Nat. Struct. Biol.* **8** 770–4

[31]     Westergard L, Christensen H M and Harris D A 2007 The cellular prion protein (PrPC): Its physiological function and role in disease *Biochim. Biophys. Acta - Mol. Basis Dis.* **1772** 629–44

[32]     Provenzano L, Ryan Y, Hilton D A, Lyons-Rimmer J, Dave F, Maze E A, Adams C L, Rigby-Jones R, Ammoun S and Hanemann C O 2017 Cellular prion protein (PrP C) in the development of Merlin-deficient tumours *Oncogene* **36** 6132–42

[33]     Linsenmeier L, Altmeppen H C, Wetzel S, Mohammadi B, Saftig P and Glatzel M 2017 Diverse functions of the prion protein – Does proteolytic processing hold the key? *Biochim. Biophys. Acta - Mol. Cell Res.* **1864** 2128–37

[34]     Wu G R, Mu T C, Gao Z X, Wang J, Sy M S and Li C Y 2017 Prion protein is required for tumor necrosis factor α (TNFα)-triggered nuclear factor κb (NF-κB) signaling and cytokine production *J. Biol. Chem.* **292** 18747–59

[35]     Linden R 2017 The biological function of the prion protein: A cell surface scaffold of signaling modules *Front. Mol. Neurosci.* **10** 77

[36]     Franzmann T M, Jahnel M, Pozniakovsky A, Mahamid J, Holehouse A S, Nüske E, Richter D, Baumeister W, Grill S W, Pappu R V., Hyman A A and Alberti S 2018 Phase separation of a yeast prion protein promotes cellular fitness *Science (80-. ).* **359** eaao5654

[37]     Pan K M, Baldwin M, Nguyen J, Gasset M, Serban A, Groth D, Mehlhorn I, Huang Z, Fletterick R J, Cohen F E and Prusiner S B 1993 Conversion of α-helices into β-sheets features in the formation of the scrapie prion proteins *Proc. Natl. Acad. Sci. U. S. A.* **90** 10962–6

[38]     Diaz-Espinoza R and Soto C 2012 High-resolution structure of infectious prion protein: The final frontier *Nat. Struct. Mol. Biol.* **19** 370–7

[39]     Wille H and Requena J 2018 The Structure of PrPSc Prions *Pathogens* **7** 20

[40]   Zou W-Q, Zhou X, Yuan J and Xiao X 2011 Insoluble cellular prion protein and its association with prion and Alzheimer diseases *Prion* **5** 172–8

[41]   Ma J and Lindquist S 2002 Conversion of PrP to a self-perpetuating PrPSc-like conformation in the cytosol *Science (80-. ).* **298** 1785–8

[42]   Jackson G S 1999 Reversible conversion of monomeric human prion protein between native and fibrilogenic conformations *Science (80-. ).* **283** 1935–7

[43]   Wang H, Rhoads D D and Appleby B S 2019 Human prion diseases *Curr. Opin. Infect. Dis.* **32** 272–6

[44]   Petersen R B, Parchi P, Richardson S L, Urig C B and Gambetti P 1996 Effect of the D178N mutation and the codon 129 polymorphism on the metabolism of the prion protein *J. Biol. Chem.* **271** 12661–8

[45]   Zarranz J J, Digon A, Atarés B, Rodríguez-Martinez A B, Arce A, Carrera N, Fernández-Manchola I, Fernández-Martínez M, Fernández-Maiztegui C, Forcadas I, Galdos L, Gómez-Esteban J C, Ibáñez A, Lezcano E, De López Munain A, Martí-Massó J F, Mendibe M M, Urtasun M, Uterga J M, Saracibar N, Velasco F and De Pancorbo M M 2005 Phenotypic variability in familial prion diseases due to the D178N mutation *J. Neurol. Neurosurg. Psychiatry* **76** 1491–6

[46]   Woulfe J, Kertesz A, Frohn I, Bauer S, George-Hyslop P S and Bergeron C 2005 Gerstmann-Sträussler-Scheinker disease with the Q217R mutation mimicking frontotemporal dementia *Acta Neuropathol.* **110** 317–9

[47]   Tartaglia M C, Thai J N, See T, Kuo A, Harbaugh R, Raudabaugh B, Cali I, Sattavat M, Sanchez H, DeArmond S J and Geschwind M D 2010 Pathologic Evidence That the T188R Mutation in PRNP Is Associated With Prion Disease *J. Neuropathol. Exp. Neurol.* **69** 1220–7

[48]   Palmer M S, Dryden A J, Hughes J T and Collinge J 1991 Homozygous prion protein genotype predisposes to sporadic Creutzfeldt-Jakob disease *Nature* **352** 340–2

[49]   Shibuya S, Higuchi J, Shin R W, Tateishi J and Kitamoto T 1998 Codon 219 Lys allele of PRNP is not found in sporadic Creutzfeldt-Jakob disease *Ann. Neurol.* **43** 826–8

[50]   Soldevila M, Calafell F, Andrés A M, Yagüe J, Helgason A, Stefánsson K and Bertranpetit J 2003 Prion susceptibility and protective alleles exhibit marked geographic differences *Hum. Mutat.* **22** 104–5

[51]   Takayanagi M, Suzuki K, Nakamura T, Hirata K, Satoh K and Kitamoto T 2018 Genetic Creutzfeldt-Jakob disease with a glutamate-to-lysine substitution at codon 219 (E219K) in the presence of the E200K mutation presenting with rapid progressive dementia following slowly progressive clinical course *Rinsho Shinkeigaku* **58** 682–7

[52]   Mead S, Whitfield J, Poulter M, Shah P, Uphill J, Campbell T, Al-Dujaily H, Hummerich H, Beck J, Mein C A, Verzilli C, Whittaker J, Alpers M P and Collinge J 2009 A novel protective prion protein variant that colocalizes with kuru exposure *N. Engl. J. Med.* **361** 2056–65

[53]   Asante E A, Smidak M, Grimshaw A, Houghton R, Tomlinson A, Jeelani A, Jakubcova T, Hamdan S, Richard-Londt A, Linehan J M, Brandner S, Alpers M, Whitfield J, Mead S, Wadsworth J D F and Collinge J 2015 A naturally occurring variant of the human prion protein completely prevents prion disease *Nature* **522** 478–81

[54]   Sabareesan A T and Udgaonkar J B 2017 The G126V Mutation in the Mouse Prion Protein Hinders Nucleation-Dependent Fibril Formation by Slowing Initial Fibril Growth and by Increasing the Critical Concentration *Biochemistry* **56** 5931–42

[55] Pan A C, Jacobson D, Yatsenko K, Sritharan D, Weinreich T M and Shaw D E 2019 Atomic-level characterization of protein–protein association *Proc. Natl. Acad. Sci. U. S. A.* **116** 4244–9

[56] Bermudez M, Mortier J, Rakers C, Sydow D and Wolber G 2016 More than a look into a crystal ball: protein structure elucidation guided by molecular dynamics simulations *Drug Discov. Today* **21** 1799–805

[57] Plattner N, Doerr S, De Fabritiis G and Noé F 2017 Complete protein-protein association kinetics in atomic detail revealed by molecular dynamics simulations and Markov modelling *Nat. Chem.* **9** 1005–11

[58] Zhou S, Shi D, Liu X, Liu H and Yao X 2016 Protective V127 prion variant prevents prion disease by interrupting the formation of dimer and fibril from molecular dynamics simulations *Sci. Rep.* **6** 1–12

[59] Zheng Z, Zhang M, Wang Y, Ma R, Guo C, Feng L, Wu J, Yao H and Lin D 2018 Structural basis for the complete resistance of the human prion protein mutant G127V to prion disease *Sci. Rep.* **8** 1–15

[60] El-Bastawissy E, Knaggs M H and Gilbert I H 2001 Molecular dynamics simulations of wild-type and point mutation human prion protein at normal and elevated temperature *J. Mol. Graph. Model.* **20** 145–54

[61] Sekijima M, Motono C, Yamasaki S, Kaneko K and Akiyama Y 2003 Molecular dynamics simulation of dimeric and monomeric forms of human prion protein: Insight into dynamics and properties *Biophys. J.* **85** 1176–85

[62] Mandujano-Rosas L A, Osorio-González D, Reyes-Romero P G and Mulia-Rodríguez J 2014 Human Prion Protein Conformational Changes Susceptibility: A Molecular Dynamics Simulation Study *Open J. Biophys.* **04** 169–75

[63] Borgohain G, Dan N and Paul S 2016 Use of molecular dynamics simulation to explore structural facets of human prion protein with pathogenic mutations *Biophys. Chem.* **213** 32–9

[64] Gao Y, Zhu T, Zhang C, Zhang J Z H and Mei Y 2018 Comparison of the unfolding and oligomerization of human prion protein under acidic and neutral environments by molecular dynamics simulations *Chem. Phys. Lett.* **706** 594–600

[65] Barducci A, Chelli R, Procacci P, Schettino V, Gervasio F L and Parrinello M 2006 Metadynamics simulation of prion protein: β-structure stability and the early stages of misfolding *J. Am. Chem. Soc.* **128** 2705–10

[66] Caldarulo E, Barducci A, Wüthrich K and Parrinello M 2017 Prion protein β2–α2 loop conformational landscape *Proc. Natl. Acad. Sci. U. S. A.* **114** 9617–22

[67] Huang D and Caflisch A 2015 Evolutionary conserved Tyr169 stabilizes the β2-α2 loop of the prion protein *J. Am. Chem. Soc.* **137** 2948–57

[68] Avbelj M, Hafner-Bratkovič I and Jerala R 2011 Introduction of glutamines into the B2-H2 loop promotes prion protein conversion *Biochem. Biophys. Res. Commun.* **413** 521–6

[69] Kurt T D, Aguilar-Calvo P, Jiang L, Rodriguez J A, Alderson N, Eisenberg D S and Sigurdson C J 2017 Asparagine and glutamine ladders promote cross-species prion conversion *J. Biol. Chem.* **292** 19076–86

[70] Sigurdson C J, Nilsson K P R, Hornemann S, Manco G, Fernández-Borges N, Schwarz P, Castilla J, Wüthrich K and Aguzzi A 2010 A molecular switch controls interspecies prion disease transmission in mice *J. Clin. Invest.* **120** 2590–9

[71]     Christen B, Pérez D R, Hornemann S and Wüthrich K 2008 NMR Structure of the Bank Vole Prion Protein at 20 °C Contains a Structured Loop of Residues 165-171 *J. Mol. Biol.* **383** 306–12

[72]     Gossert A D, Bonjour S, Lysek D A, Fiorito F and Wüthrich K 2005 Prion protein NMR structures of elk and of mouse/elk hybrids *Proc. Natl. Acad. Sci. U. S. A.* **102** 646–50

[73]     Sweeting B, Brown E, Khan M Q, Chakrabartty A and Pai E F 2013 N-Terminal Helix-Cap in α-Helix 2 Modulates β-State Misfolding in Rabbit and Hamster Prion Proteins *PLoS One* **8** e63047

[74]     Pérez D R, Damberger F F and Wüthrich K 2010 Horse Prion Protein NMR Structure and Comparisons with Related Variants of the Mouse Prion Protein *J. Mol. Biol.* **400** 121–8

[75]     Agarwal S, Döring K, Gierusz L A, Iyer P, Lane F M, Graham J F, Goldmann W, Pinheiro T J T and Gill A C 2015 Complex folding and misfolding effects of deer-specific amino acid substitutions in the β2-α2 loop of murine prion protein *Sci. Rep.* **5** 1–14

[76]     Kaneko K, Zulianello L, Scott M, Cooper C M, Wallace A C, James T L, Cohen F E and Prusiner S B 1997 Evidence for protein X binding to a discontinuous epitope on the cellular prion protein during scrapie prion propagation *Proc. Natl. Acad. Sci. U. S. A.* **94** 10069–74

[77]     Knaus K J, Morillas M, Swietnicki W, Malone M, Surewicz W K and Yee V C 2001 Crystal structure of the human prion protein reveals a mechanism for oligomerization *Nat. Struct. Biol.* **8** 770–4

[78]     Bjorndahl T C, Zhou G P, Liu X, Perez-Pineiro R, Semenchenko V, Saleem F, Acharya S, Bujold A, Sobsey C A and Wishart D S 2011 Detailed biophysical characterization of the acid-induced PrPc to PrPβ conversion process *Biochemistry* **50** 1162–73

[79]     Singh J, Kumar H, Sabareesan A T and Udgaonkar J B 2014 Rational stabilization of helix 2 of the prion protein prevents its misfolding and oligomerization *J. Am. Chem. Soc.* **136** 16704–7

[80]     Gower J C and Ross G J S 1969 Minimum Spanning Trees and Single Linkage Cluster Analysis *Appl. Stat.* **18** 54

[81]     Huang J J, Li X N, Liu W L, Yuan H Y, Gao Y, Wang K, Tang B, Pang D W, Chen J and Liang Y 2020 Neutralizing Mutations Significantly Inhibit Amyloid Formation by Human Prion Protein and Decrease Its Cytotoxicity *J. Mol. Biol.* **432** 828–44

[82]     Zuegg J and Gready J E 1999 Molecular dynamics simulations of human prion protein: Importance of correct treatment of electrostatic interactions *Biochemistry* **38** 13862–76

[83]     Zhang J 2011 Comparison studies of the structural stability of rabbit prion protein with human and mouse prion proteins *J. Theor. Biol.* **269** 88–95

[84]     Miao Y, Nichols S E, Gasper P M, Metzger V T and McCammon J A 2013 Activation and dynamic network of the M2 muscarinic receptor *Proc. Natl. Acad. Sci. U. S. A.* **110** 10982–7

[85]     del Sol A, Tsai C J, Ma B and Nussinov R 2009 The Origin of Allosteric Functional Modulation: Multiple Pre-existing Pathways *Structure* **17** 1042–50

[86]     Csermely P, Palotai R and Nussinov R 2010 Induced fit, conformational selection and independent dynamic segments: an extended view of binding events *Nat. Preced.* 1–9

[87]     Csermely P, Korcsmáros T, Kiss H J M, London G and Nussinov R 2013 Structure and dynamics of molecular networks: A novel paradigm of drug discovery: A comprehensive review *Pharmacol. Ther.* **138** 333–408

[88]     Sethi A, Eargle J, Black A A and Luthey-Schulten Z 2009 Dynamical networks in tRNA: Protein complexes *Proc. Natl. Acad. Sci. U. S. A.* **106** 6620–5

[89]     Jasper J. Koehorst, Jesse C. J. van Dam, Edoardo Saccenti, Vitor A. P. Martins dos Santos M S-D and P J S 2017 GISAID Global Initiative on Sharing All Influenza Data. Phylogeny of SARS-like betacoronaviruses including novel coronavirus (nCoV) *Oxford* **34** 1401–3

[90]     Ahalawat N and Murarka R K 2015 Conformational changes and allosteric communications in human serum albumin due to ligand binding *J. Biomol. Struct. Dyn.* **33** 2192–204

[91]     Yang J, Liu H, Liu X, Gu C, Luo R and Chen H F 2016 Synergistic Allosteric Mechanism of Fructose-1,6-bisphosphate and Serine for Pyruvate Kinase M2 via Dynamics Fluctuation Network Analysis *J. Chem. Inf. Model.* **56** 1184–92

[92]     Alcalá-Corona S A, Velázquez-Caldelas T E, Espinal-Enríquez J and Hernández-Lemus E 2016 Community Structure Reveals Biologically Functional Modules in MEF2C Transcriptional Regulatory Network *Front. Physiol.* **7** 184

[93]     Guo C and Zhou H X 2016 Unidirectional allostery in the regulatory subunit RIα facilitates efficient deactivation of protein kinase A *Proc. Natl. Acad. Sci. U. S. A.* **113** E6776–85

[94]     Girvan M and Newman M E J 2002 Community structure in social and biological networks *Proc. Natl. Acad. Sci. U. S. A.* **99** 7821–6

[95]     for L S-G S T is no corresponding record The PyMOL Molecular Graphics System, Version 2.0 Schrödinger, LLC (2017)

[96]     Lindorff-Larsen K, Piana S, Palmo K, Maragakis P, Klepeis J L, Dror R O and Shaw D E 2010 Improved side-chain torsion potentials for the Amber ff99SB protein force field *Proteins Struct. Funct. Bioinforma.* **78** 1950–8

[97]     Abraham M J, Murtola T, Schulz R, Páll S, Smith J C, Hess B and Lindah E 2015 Gromacs: High performance molecular simulations through multi-level parallelism from laptops to supercomputers *SoftwareX* **1–2** 19–25

[98]     Darden T, York D and Pedersen L 1993 Particle mesh Ewald: An N·log(N) method for Ewald sums in large systems *J. Chem. Phys.* **98** 10089–92

[99]     Bussi G, Donadio D and Parrinello M 2007 Canonical sampling through velocity rescaling *J. Chem. Phys.* **126** 014101

[100]    Parrinello M and Rahman A 1981 Polymorphic transitions in single crystals: A new molecular dynamics method *J. Appl. Phys.* **52** 7182–90

[101]    Páll S and Hess B 2013 A flexible algorithm for calculating pair interactions on SIMD architectures *Comput. Phys. Commun.* **184** 2641–50

[102]    Papadimitriou C and Sideri M 1999 On the Floyd-Warshall algorithm for logic programs *J. Log. Program.* **41** 129–37

[103]    Newman M E J and Girvan M 2004 Finding and evaluating community structure in networks *Phys. Rev. E - Stat. Nonlinear, Soft Matter Phys.* **69** 1–15

[104]    Humphrey W, Dalke A and Schulten K 1996 VMD: Visual molecular dynamics *J. Mol. Graph.* **14** 33–8

# Supplementary Information

# Structural and Dynamical Mechanisms of a Naturally Occurring Variant of the Human Prion Protein in Preventing Prion Conversion

Yiming Tang, Yifei Yao, and Guanghong Wei*

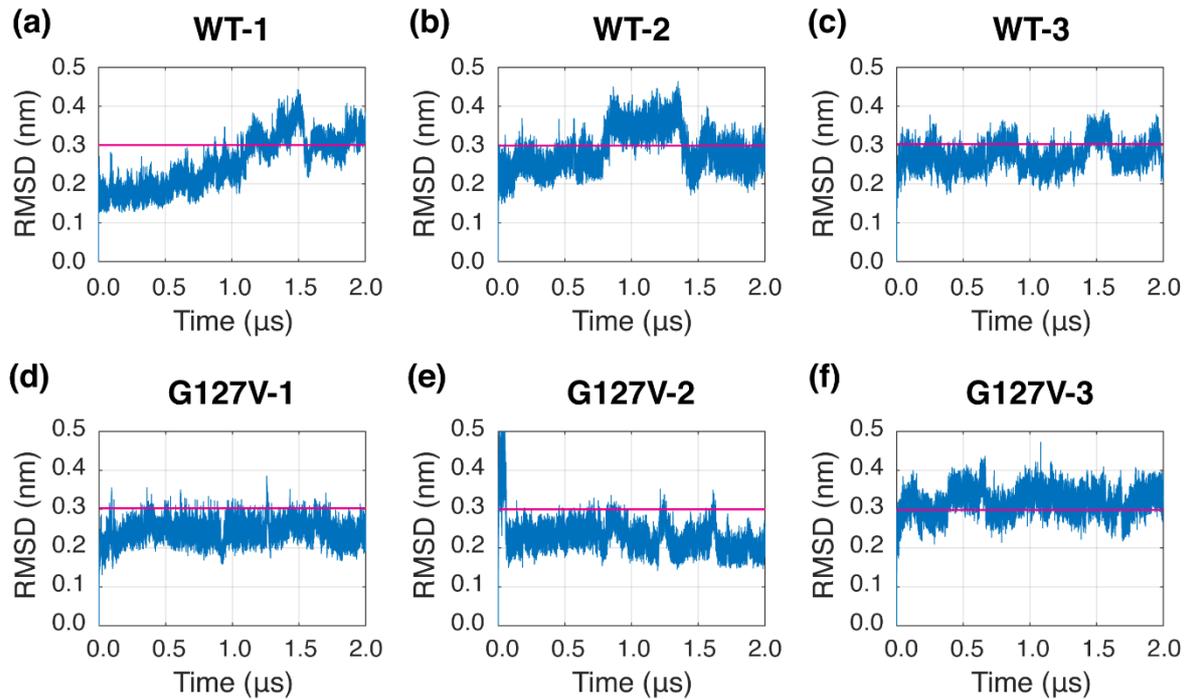

**Figure S1.** Time evolution of RMSD values with respect to the native NMR structure of WT PrP in (a-c) three WT systems, and (d-f) three G127V systems.

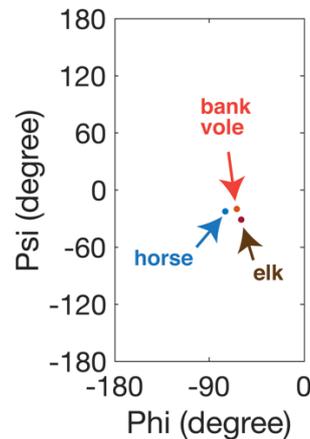

**Figure S2.** The Ramachandran plot of residue M166 in bank vole, elk and horse PrPs.

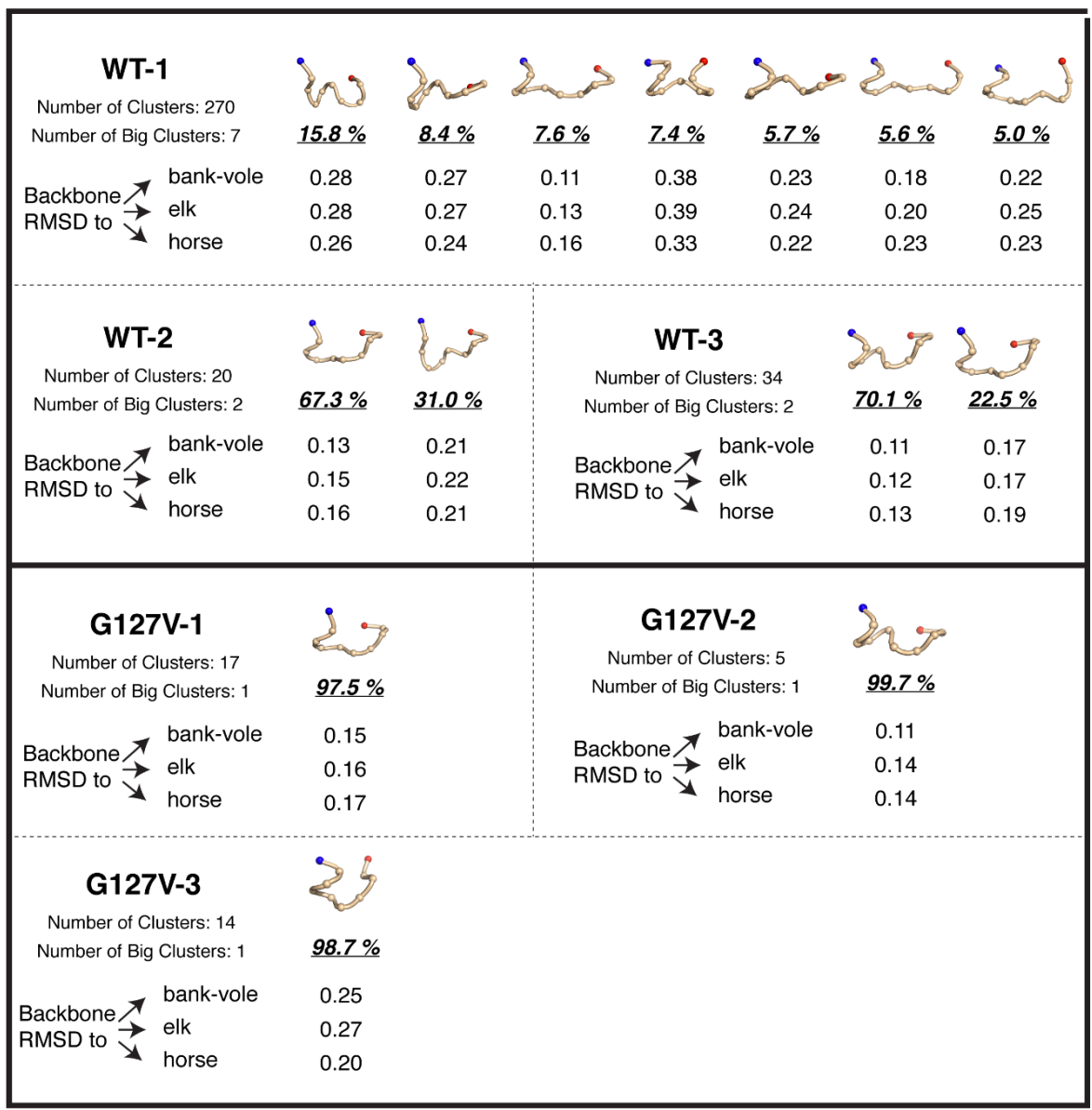

**Figure S3. Cluster analysis of the S2-H2 loop conformations.** Snapshots of the representative configurations are shown with the numbers under them corresponding to their proportions. The backbone RMSD of representative structure of each cluster to the S2-H2 loop in three rigid-loop PrPs (bank-vole, elk, and horse PrP) are also shown.

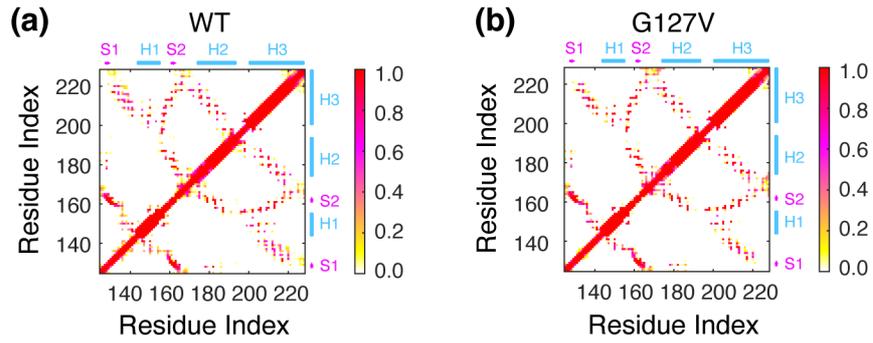

**Figure S4.** Adjacency matrices of (a) WT and (b) G127V system.

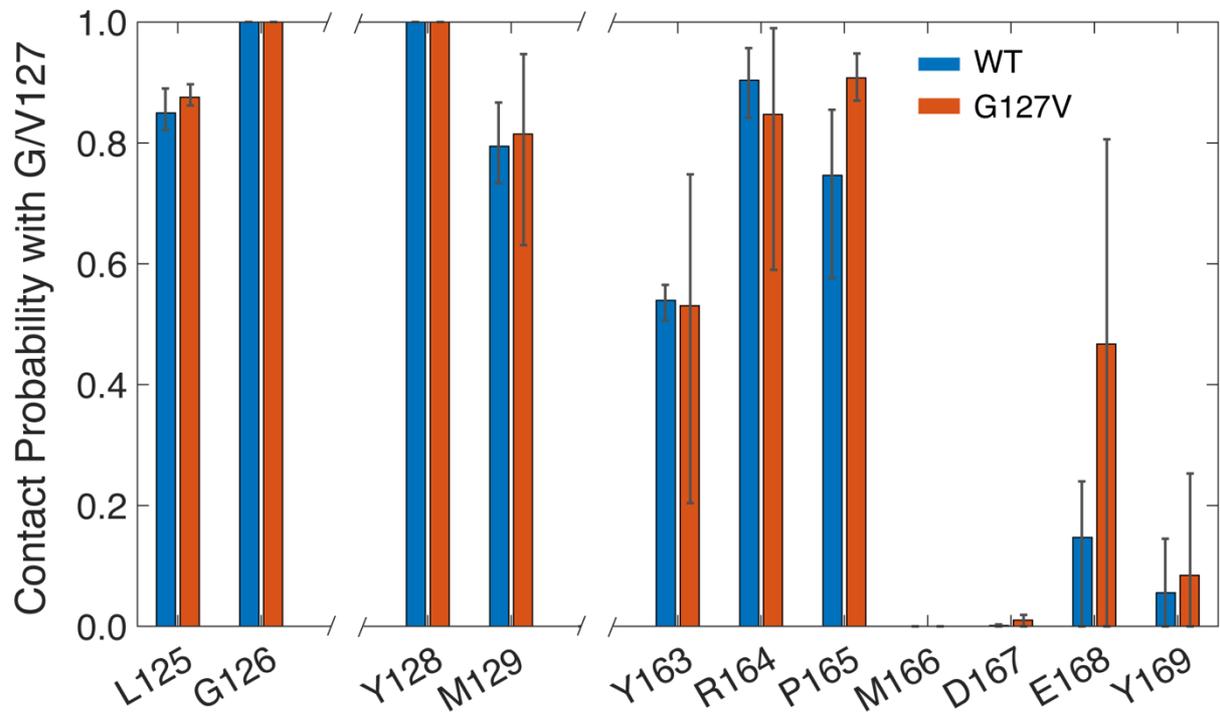

**Figure S5.** Contact probabilities of G/V127 with (1) all residues whose contact probability with G/V127 is larger than 0.5, and (2) all residues on the S2-H2 loop.

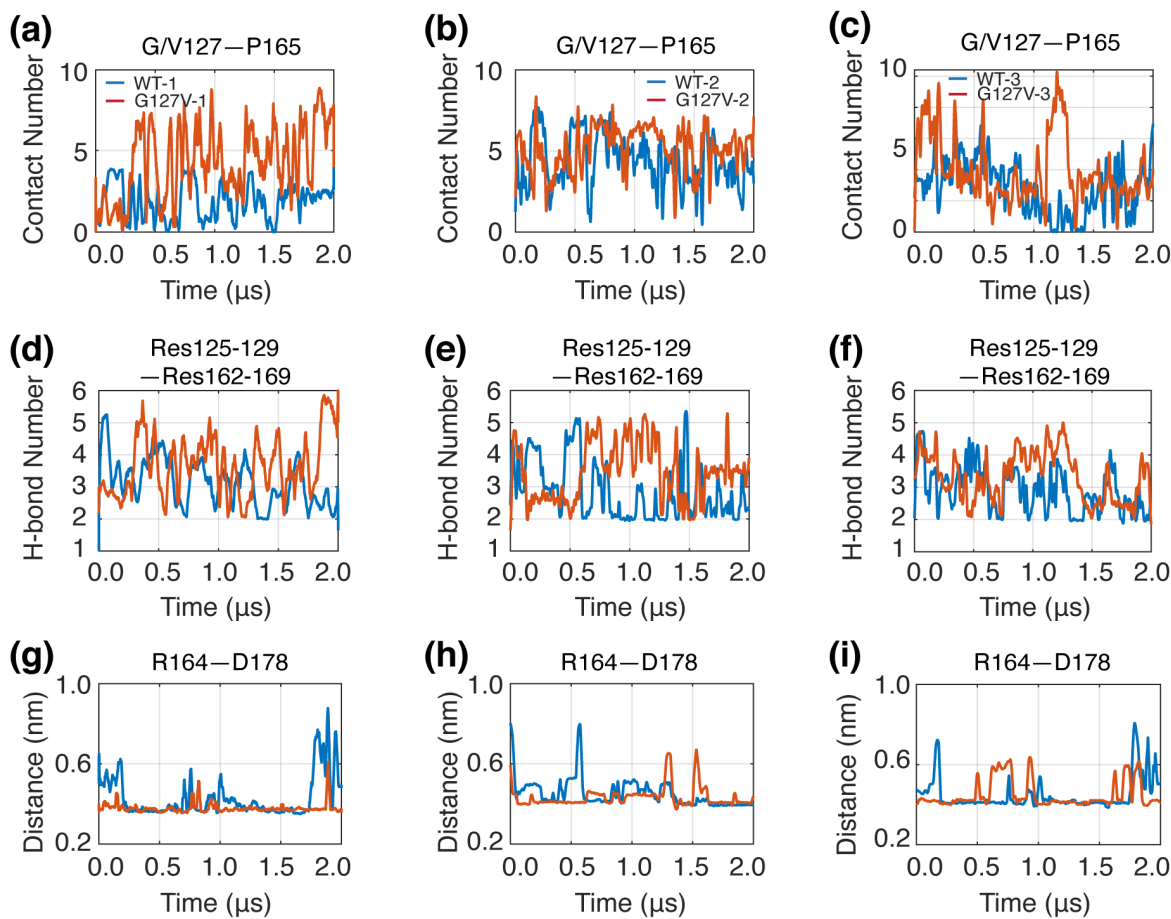

**Figure S6.** Influence of G127V mutation on the interactions in the vicinity of residue 127 and S2-H2 loop. (a, b, c) Time evolution of the contact number between G/V127 and P165. (d, e, f) Time evolution of the number of H-bonds between residues 125~129 and residues 162~169. (g, h, i) Time evolution of the centroid distance between residue R164 and D178 charged sidechain groups.

Table S1. Optimal allosteric path from mutation site (G127) to G195 in WT systems.

| Ranking | Length | HEAD | Nodes on the Path | | | | | | END |
|---|---|---|---|---|---|---|---|---|---|
| | | | Nodes in S1 | Node in S2 | (C-terminal) | Nodes in H3 | (N-terminal) | H2-H3 Loop | |
| 1 | 317 | G127 | M129 | V161 | | V210    R208 | M205  D202 | T199  N197 | G195 |
| 1 | 317 | G127 | M129 | V161 | | V210 | M206    D202 | T199  N197 | G195 |
| 3 | 321 | G127 | M129 | V161 | C214    E211 | R208 | M205  D202 | T199  N197 | G195 |
| 4 | 322 | G127 | M129  G131 | V161 | | V210    R208 | M205  D202 | T199  N197 | G195 |
| 4 | 322 | G127 | M129  G131 | V161 | | V210 | M206    D202 | T199  N197 | G195 |
| 6 | 323 | G127 | M129 | V161 | M213 | V210    R208 | M205  D202 | T199  N197 | G195 |
| 6 | 323 | G127 | M129 | V161 | M213 | V209 | M205  D202 | T199  N197 | G195 |
| 6 | 323 | G127 | M129 | V161 | M213 | V210 | M206    D202 | T199  N197 | G195 |
| 9 | 325 | G127 | M129 | V161 | C214    Q212 | V209 | M205  D202 | T199  N197 | G195 |
| 10 | 326 | G127 | M129 | V161 | C214 | V210    R208 | M205  D202 | T199  N197 | G195 |
| 10 | 326 | G127 | M129  G131 | V161 | C214    E211 | R208 | M205  D202 | T199  N197 | G195 |
| 10 | 326 | G127 | M129 | V161 | M213    E211 | R208 | M205  D202 | T199  N197 | G195 |
| 10 | 326 | G127 | M129 | V161 | | V210    E207 | M205  D202 | T199  N197 | G195 |
| 10 | 326 | G127 | M129 | V161 | C214 | V210 | M206    D202 | T199  N197 | G195 |
| 10 | 326 | G127 | M129 | V161 | M213 | V209 | M206    D202 | T199  N197 | G195 |
| 16 | 327 | G127 | M129 | V161 | C214    Q212 | R208 | M205  D202 | T199  N197 | G195 |
| 16 | 327 | G127 | M129 | V161 | | V210    R208 | M206    D202 | T199  N197 | G195 |

Table S2. Optimal allosteric path from mutation site (V127) to G195 in G127V systems.

| Ranking | Length | HEAD | Nodes on the Path |||||||| END |
|---|---|---|---|---|---|---|---|---|---|---|---|
| | | | Nodes in S1 | Nodes in S2 | (N-terminal) | | Nodes in H2 | | (C-terminal) | | |
| 1 | 257 | V127 | | M129 | Y162 | I182 | K185 | T188 | | T192 | | G195 |
| 1 | 257 | V127 | | M129 | Y162 | C179 | I182 | K185 | T188 | | T192 | | G195 |
| 3 | 258 | V127 | | M129 | Y162 | | I182 | K185 | | V189 | | T193 | G195 |
| 3 | 258 | V127 | | M129 | Y162 | C179 | I182 | K185 | | V189 | | T193 | G195 |
| 3 | 258 | V127 | | M129 | Y162 | | | | Q186 | V189 | | T193 | G195 |
| 3 | 258 | V127 | | M129 | Y162 | | T183 | | Q186 | V189 | | T193 | G195 |
| 7 | 260 | V127 | | M129 | Y162 | C179 | T183 | | Q186 | V189 | | T193 | G195 |
| 8 | 262 | V127 | | M129 | Y162 | | I182 | | Q186 | V189 | | T193 | G195 |
| 8 | 262 | V127 | | M129 | Y162 | C179 | I182 | | Q186 | V189 | | T193 | G195 |
| 10 | 263 | V127 | | M129 | Y162 | | I182 | K185 | | V189 | T192 | | G195 |
| 10 | 263 | V127 | | M129 | Y162 | C179 | I182 | K185 | | V189 | T192 | | G195 |
| 10 | 263 | V127 | | M129 | Y162 | | | | Q186 | V189 | T192 | | G195 |
| 10 | 263 | V127 | | M129 | Y162 | | T183 | | Q186 | V189 | T192 | | G195 |
| 14 | 264 | V127 | | M129 | Y162 | | T183 | K185 | T188 | | T192 | | G195 |
| 15 | 265 | V127 | R164 | Y128 | Y162 | C179 | I182 | K185 | T188 | | T192 | | G195 |
| 15 | 265 | V127 | | M129 | Y162 | C179 | T183 | | Q186 | V189 | T192 | | G195 |
| 15 | 265 | V127 | | M129 | Y162 | | T183 | K185 | | V189 | | T193 | G195 |
| 18 | 266 | V127 | | M129 | Y162 | C179 | T183 | K185 | T188 | | T192 | | G195 |
| 18 | 266 | V127 | | M129 | Y162 | | | | Q186 | T188 | | T192 | | G195 |
| 18 | 266 | V127 | | M129 | Y162 | | T183 | | Q186 | T188 | | T192 | | G195 |
| 18 | 266 | V127 | R164 | Y128 | Y162 | C179 | I182 | K185 | | V189 | | T193 | G195 |
| 18 | 266 | V127 | R164 | Y128 | Y162 | | T183 | | Q186 | V189 | | T193 | G195 |
| 23 | 267 | V127 | | M129 | Y162 | | I182 | | Q186 | V189 | T192 | | G195 |
| 23 | 267 | V127 | | M129 | Y162 | C179 | I182 | | Q186 | V189 | T192 | | G195 |
| 23 | 267 | V127 | | M129 | Y162 | C179 | T183 | K185 | | V189 | | T193 | G195 |

Table S3. Suboptimal allosteric path from mutation site (V127) to G195 in G127V systems.

| Ranking | Length | HEAD | Nodes on the Path | | | | | | | | END |
|---|---|---|---|---|---|---|---|---|---|---|---|
| | | | S1 | S2 | (C-terminal) <------ | | Nodes in H3 | ------> (N-terminal) | H2-H3 Loop | | |
| 1 | 282 | V127 | M129 | V161 | | V210 | E207 | K204 | T201 | T199 | N197 | G195 |
| 2 | 284 | V127 | M129 | V161 | C214 | E211 | E207 | K204 | T201 | T199 | N197 | G195 |
| 3 | 288 | V127 | M129 | V161 | C214 | V210 | E207 | K204 | T201 | T199 | N197 | G195 |
| 4 | 290 | V127 | M129 | V161 | | V210 | E207 | K204 | D202 | T199 | N197 | G195 |
| 5 | 292 | V127 | M129 | V161 | C214 | E211 | E207 | K204 | D202 | T199 | N197 | G195 |
| 5 | 292 | V127 | M129 | V161 | | V210 | E207 | V203 | | T199 | N197 | G195 |
| 5 | 292 | V127 | M129 | V161 | | V210 | E207 | K204 | D202 | | N197 | G195 |